\documentclass{jpsj3}
\usepackage{txfonts}
\usepackage{epstopdf, epsfig}
\usepackage{xcolor}
\usepackage{color,bm,ascmac,amssymb,amsmath}
\newcommand{\average}[1]{\ensuremath{\langle#1\rangle} }

\usepackage{psfrag}

\title{Data-driven low-dimensional model for the onset of turbulence \\
in plane Couette flow}

\author{Masaki Shimizu$^1$\thanks{shimizu@me.es.osaka-u.ac.jp} 
,  Julius Rhoan T. Lustro$^2$  and Genta Kawahara$^1$}
\inst{$^1$Graduate School of Engineering Science, the University of Osaka, 560-8531 Japan \\
$^2$Fluids Laboratory, Department of Mechanical Engineering, University of the Philippines Diliman, Quezon City, 1101 Philippines} %\\

\abst{We construct low-dimensional dynamical systems for low-Reynolds-number turbulence in plane Couette flow using Kernel Quantile Regression. Exploiting the finite-dimensional structure of attractors in dissipative systems, reduced-order models are learned from direct numerical simulation data using a small set of physically meaningful observables. The resulting discrete-time models accurately reproduce periodic and chaotic dynamics near the onset of turbulence. The minimum number of variables required for accurate prediction is found to be consistent with embedding theory based on the attractor dimension. For chaotic regimes, the models capture both short-term trajectory evolution and long-term statistical properties, including probability density functions. By incorporating the Reynolds number as an additional input variable, we further develop a parameter-dependent model that successfully reproduces the bifurcation structure over a range of Reynolds numbers, including bifurcations between training points. These results demonstrate the effectiveness of machine-learning-based reduced-order modeling for capturing the essential dynamics and statistics of weakly turbulent flows. }

\begin{document}
\maketitle

\section{Introduction}

The asymptotic behavior of a dissipative system is often confined to an attractor whose dimension is much lower than that of the original phase space\cite{temam2012infinite}. If the Hausdorff dimension of an attractor $A$ is $N_D$, then almost every projection into a space of dimension $2\lfloor N_D \rfloor + 1$ is one-to-one on $A$, where $\lfloor \cdot \rfloor$ denotes the floor function\cite{sauer1991embedology}. 
Consequently, when the attractor effectively occupies only a small number of independent directions, the system state can, in principle, be described by a relatively small set of variables. This implies that a low-dimensional model can accurately reproduce the dynamics of the original system.

Data-driven methods for identifying dynamical systems from observational data can be traced back to the pioneering work of Nagumo in control theory \cite{nagumo1967learning}. Recent advances in machine learning algorithms, together with increasing computational power, have accelerated the development of dimensionality reduction techniques that leverage the sparse structure of data. As a result, these methods have found widespread applications across a variety of disciplines, including fluid dynamics.
One of the most successful data-driven approaches is dynamic mode decomposition (DMD) \cite{schmid2010dynamic,rowley2009spectral}. Its success is largely attributed to its close connection with Koopman spectral theory \cite{mezic2013analysis,kutz2016dynamic}. DMD decomposes flow fields into simple spatio-temporal coherent structures, enabling accurate data compression and providing valuable insights into the underlying dynamics.
In parallel, artificial neural network-based statistical learning methods, which have fueled the recent boom in artificial intelligence, have been applied to improve turbulence modeling in both Reynolds-averaged Navier-Stokes (RANS) simulations \cite{tracey2015machine,ling2016reynolds,wang2017physics,wu2018physics} and large-eddy simulations (LES) \cite{gamahara2017searching,maulik2017neural,beck2018neural}. Reservoir computing has also been shown to successfully reproduce the dynamics of several dissipative systems \cite{pathak2017using,pathak2018model}.
 
The success of many learning procedures is often attributed to the sparsity of the underlying data. In the case of viscous fluid flows, the theory of dissipative dynamical systems guarantees the existence of a sparse structure on the attractor. In this study, we consider the onset of chaos in plane Couette flow within a minimal periodic domain \cite{kreilos2012periodic} as a representative example. In this system, a period-doubling cascade from the steady Nagata solution
\cite{nagata1990three} leads to chaos in a manner analogous to that of a unimodal map, suggesting that the dynamics can be parameterized by only a few variables. Using the machine learning method known as kernel quantile regression \cite{takeuchi2006nonparametric}, we construct low-dimensional models based on a small set of physically meaningful quantities that accurately reproduce the original dynamics. We then 
evaluate the accuracy with which the learning process can reconstruct low-dimensional systems whose existence is predicted by underlying theory. Furthermore, by including the Reynolds number as an input feature, we demonstrate that the resulting model also reproduces the bifurcation structure of the invariant solutions responsible for the emergence of weak turbulent dynamics.

\section{Construction method of a low-dimensional system}

Consider plane Couette flow. The non-dimensional governing equations, namely the incompressible Navier--Stokes equations, together with the boundary conditions, are given by
\begin{align}
& \nabla \cdot \bm u =0,  \label{divu} \\
& \frac{\partial \bm u}{\partial t}   = \bm u \times \bm \omega -\nabla p 
+ \frac{1}{Re} \Delta \bm u,  \\
& 
{\bm u}|_{y=\pm 1}=\pm {\bm e}_x, \\
& {\bm u}(x+2\pi,y,z)={\bm u}(x,y,z+\pi)={\bm u}(x,y,z),
\end{align}
where \(x\), \(y\), and \(z\) denote the streamwise, wall-normal, and spanwise coordinates, respectively. Here, \({\bm e}_x\) is the unit vector in the \(x\)-direction, and \(Re\) is the Reynolds number based on half the wall separation and half the wall-speed difference.
The flow is restricted to the same symmetry subspace 
as that considered by Kreilos and Eckhardt,
$[u_x,u_y,u_z] (x,y,z)=[u_x,u_y,-u_z] (x+\pi,y,-z)$\cite{kreilos2012periodic}.
To compute trajectories originating from initial conditions on the attractor, the governing equations are solved by direct numerical simulation (DNS). The numerical procedure is essentially identical to that employed in Shimizu and Manneville (2019)  \cite{shimizu2018bifurcations} for channel flow, except for the wall boundary conditions and the absence of a body-force term. 
The spatial resolution is set to \((I,J,K)=(21,63,21)\), where \(2I+1\), \(J+1\), and \(2K+1\) denote the numbers of degrees of freedom in the \(x\)-, \(y\)-, and \(z\)-directions, respectively, used in the spectral expansion of spatially dependent quantities \cite{shimizu2018bifurcations}. With this resolution, the maximum errors over the computational domain in the wall-normal velocity and wall-normal vorticity fields for Nagata's solution \cite{nagata1990three} at \(Re=165\) satisfy
$\|u_y-u_y^{\rm e}\|_{\infty} < 10^{-8}$ and 
$\|\omega_y-\omega_y^{\rm e}\|_{\infty} < 10^{-6}$. 
Here, \(u_y^{\rm e}\) and \(\omega_y^{\rm e}\) denote reference solutions computed using sufficiently high spatial resolutions and are therefore regarded as exact within double-precision machine accuracy.

Here, we explain how to construct an $M$-dimensional dynamical system on an attractor.
As a training dataset, we use an orbit on the attractor obtained from a DNS.
The evolution of the orbit is then predicted using a low-dimensional system
constructed through multiple regression analysis:
\begin{equation}
X^{n+1}_i=\tilde{F}_i(X^n_1,X^n_2,\cdots,X^n_M)
\quad
(i=1,2,\cdots,M),
\label{low-dimensional_map}
\end{equation}
where $(X^n_1,X^n_2,\cdots,X^n_M)$ denotes the projection of the attractor
onto an $M$-dimensional subspace,
$\tilde{F}_i$ is the constructed $M$-dimensional mapping function,
and $n$ ($=0,1,2,\cdots$) represents the discrete time
$t=n\Delta t$.
In this study, $\Delta t=1$ in all cases.
For chaotic systems, $\Delta t$ must be sufficiently smaller than the inverse
of the maximum Lyapunov exponent to enable accurate prediction of future states.

To construct a continuous dynamical system, $ \dot{X}_i=\tilde{F}_i(X_1,X_2,\ldots,X_M)$,  it is necessary to derive evolution equations for each $\dot{X}_i$ from the Navier--Stokes equations when the time derivatives are not directly available as observables. In practice, such derivatives are often difficult to obtain from observable quantities. In contrast, for the discrete-time system considered in this study, no explicit governing equations need to be assumed. The variables $X_i$ ($i=1,2,\ldots,M$) are defined as spatially averaged quantities of physical importance, including the kinetic energy and enstrophy. Specifically, the candidate variables are \[ \frac{1}{V}\int_V \left[ u_x^2,\,u_y^2,\,u_z^2,\, \omega_x^2,\,\omega_y^2,\,\omega_z^2,\, (u_x-\langle u_x\rangle_{xz})^2,\, \Delta u_x \right]\,dV . \] The first $M$ quantities are selected to construct an $M$-dimensional system. These variables serve as predictors in the regression analysis. Provided that the variables are not strongly correlated, the predictive performance is relatively insensitive to the specific choice of variables. To model the nonlinear relationship among the variables, we employ Kernel Quantile Regression (KQR) \cite{takeuchi2006nonparametric} implemented in the Kernlab package \cite{karatzoglou2004kernlab} for R. In KQR, the fitting function is represented as a linear combination of kernel functions during training. A Gaussian kernel is used throughout this study.

We use the default values for all KQR's hyper parameters, including $\tau$ (the target quantile) and kpar, except for the regularization parameter $C$. This parameter determines the strength of regularization and therefore affects the smoothness of the fitted function. To reduce the impact of regularization, we set $C$ to a relatively large value in the range $100 \leq C \leq 1000$ \cite{takeuchi2006nonparametric,karatzoglou2004kernlab}.
In our experiments, KQR achieved the fastest reduction in prediction error as the sample size increased among all regression methods considered. One possible explanation is the difference in the underlying optimization problems: KQR is based on convex optimization, whereas artificial neural networks are typically trained through non-convex optimization procedures.

\section{Constructed low-dimensional system}

\begin{figure}%[tb]
	\begin{center}
		\psfrag{re}{$Re$}
		\psfrag{sqrt_ecf}{$E_{cf}^{1/2}$}	
		\includegraphics[width=13cm]{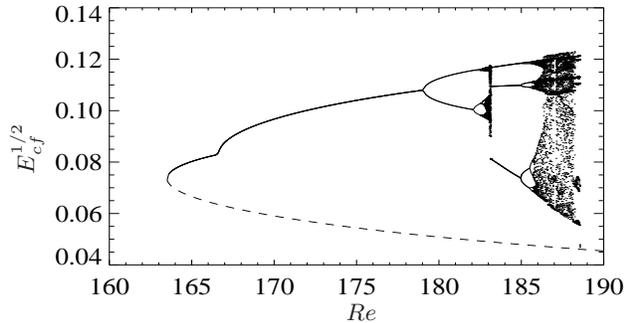}
	\end{center}
\caption{Onset of chaos %as 
due to bifurcations. 
$E_{cf}$ represents the local maximum values 
in the time sequence of the cross flow energy
$E_y+E_z$.}
\label{bifurcation}

\end{figure}

Figure \ref{bifurcation} shows the bifurcation diagram, first reported in Kreilos and Eckhardt\cite{kreilos2012periodic}, illustrating the onset of chaos (turbulence). Such bifurcation diagram, showing the instability of a coherent state—often referred to as an edge state—that leads to a boundary crisis, has been observed in low-dimensional wall-bounded shear flows\cite{avila13,zammert15,lustro19,lustro23}. For the system in this study, the Nagata steady solution emerges through a saddle-node bifurcation at $Re=163.5$\cite{nagata1990three}. 
The upper branch of the Nagata solution undergoes a bifurcation to a periodic solution, and a subsequent period-doubling cascade leads to chaotic dynamics. Figure \ref{u1u2} shows projections of the trajectories onto the $E_y$--$E_z$ plane at $Re=170$, $180$, $183$, and $183.1$, where $E_y=\average{u_y^2}_{xyz}$ and $E_z=\average{u_z^2}_{xyz}$. Here, $\average{\cdot}_{xyz}$ denotes the volume average over the periodic domain.

\subsection{Periodic case}

Since the trajectory at $Re = 170$ is periodic (Fig.~2(a)) and does not self-intersect on this projection plane, the solution curve can be parameterized by two variables. Let $E_y$ and $E_z$ be denoted by $X_1$ and $X_2$, respectively. Then, at $Re = 170$, there exist two-dimensional mapping functions $F_1$ and $F_2$ such that 
$X_1^{n+1} = F_1(X_1^n, X_2^n), X_2^{n+1} = F_2(X_1^n, X_2^n)$. 
Pairs of input vectors $(X_1^n, X_2^n)$ and their corresponding output vectors $(X_1^{n+1}, X_2^{n+1})$ are randomly sampled from a projected DNS orbit and used as the training data for the KQR algorithm. The resulting functions, denoted by $\tilde{F}_1$ and $\tilde{F}_2$, are approximations of $F_1$ and $F_2$, respectively. Their prediction accuracy is evaluated using a test set generated from a different DNS orbit than that used for training.

Figure \ref{u1t_170} (top) shows the $L_2$ error of the learned mapping as a function of the number $N$ of samples in the training set. The $L_2$ error is defined as $L_2=\left\{\overline{(\tilde{F}_1'-F_1')^2}\right\}^{1/2}$, where $\overline{(\cdot)}$ denotes the sample mean over the test set, which typically consists of a few thousand samples and is sufficiently large to evaluate statistical averages on the attractor accurately. A primed variable denotes a normalized fluctuation, defined by $f'=\frac{f-\average{f}}{\sigma},  \sigma=\average{(f-\average{f})^2}^{1/2}$, where $\average{\cdot}$ represents an average taken over a long DNS trajectory. The circles in Fig.~\ref{u1t_170} correspond to $M=2$, whereas the triangles correspond to $M=8$. The error decreases as the number of training samples increases. The asymptotic error level may depend on the regularization parameter used in the regression as well as on the temporal resolution of the DNS. Once the mappings $\tilde{F}_i$ are accurately learned, the orbit on the $E_y$--$E_z$ plane can be reproduced accurately by iterating the learned maps $\tilde{F}_i$.

Figure \ref{u1t_170} (bottom) compares the time series of the total enstrophy,
$
W=\average{\omega_x^2}_{xyz}
+\average{\omega_y^2}_{xyz}
+\average{\omega_z^2}_{xyz},
$
obtained from the Navier--Stokes trajectory (solid line), which was not included in the training set, and from the constructed two-dimensional system (red dots) with state variables $X_1=E_y$ and $X_2=E_z$.
For the constructed system, the function $W(E_y,E_z)$ is also constructed using KQR in addition to the two-dimensional dynamical system itself (see also \S 3.4). Since the system state is completely determined by the two variables $E_y$ and $E_z$, any physical quantity can be expressed as a function of these variables. The same initial conditions are used for both systems, and $N=500$ training samples are employed.

\begin{figure}%[tb]
	\begin{center}
		\psfrag{energy_y}{$E_y$}
		\psfrag{energy_z}{$E_z$}
		\psfrag{(a)}{(a)}
		\psfrag{(b)}{(b)}
		\psfrag{(c)}{(c)}
		\psfrag{(d)}{(d)}
		\includegraphics[width=12cm]{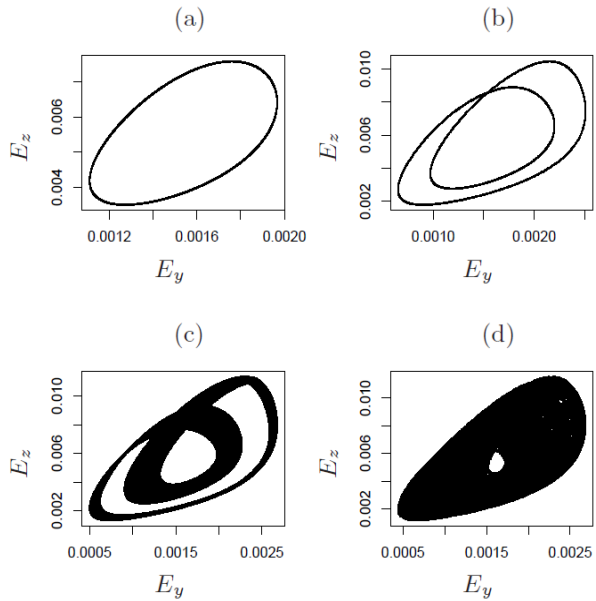}
		\caption{Projection of the orbits onto the $E_y$-$E_z$ plane
			 at $Re=$ (a) $170$, (b) $180$, (c) $183$, and (d) $183.1$.
}
		\label{u1u2}
	\end{center}
\end{figure}

\begin{figure}%[htb]
	\begin{center}
		\psfrag{L2}{$L_2$}
		\psfrag{N}{$N$}	
		\includegraphics[width=8cm,height=8cm]{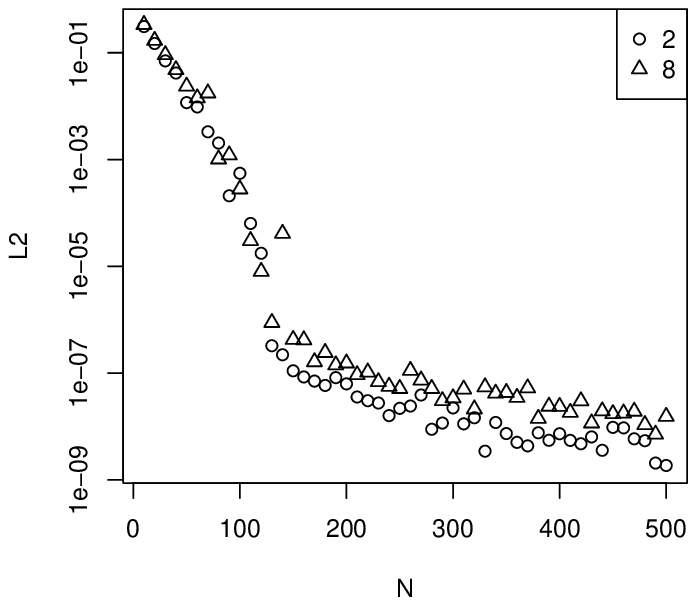}
		\psfrag{time}{$t$}
		\psfrag{enstro}{$W$}
		\includegraphics[width=8cm,height=8cm]{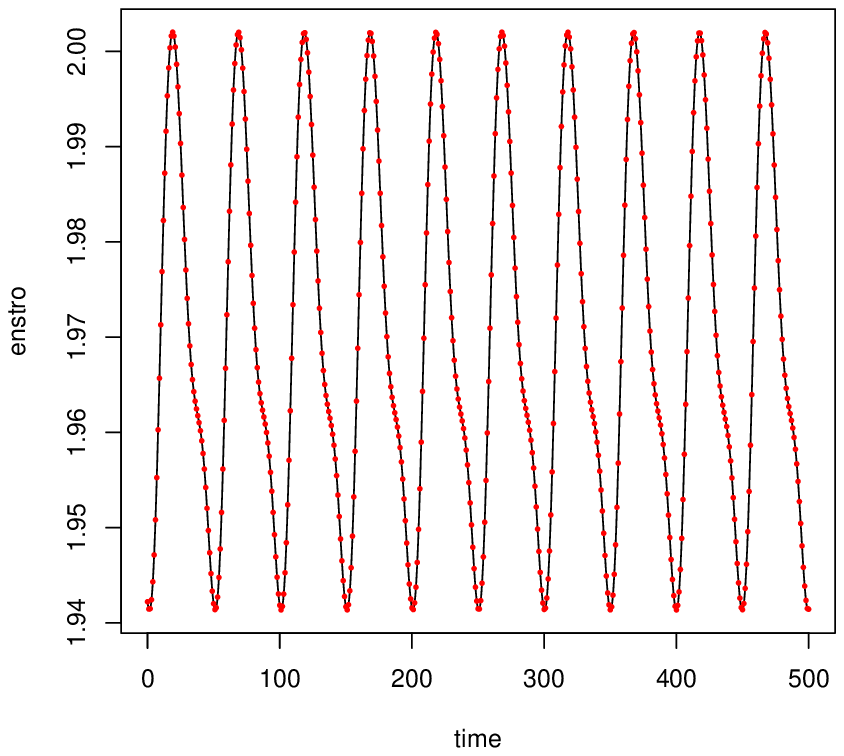}
		\caption{(Top) Dependence of model error on training set size at $Re=170$. Symbols denote the model degrees of freedom. (Bottom) Comparison of the time series of $W$  obtained from the Navier--Stokes equations (solid line) and the low-dimensional model (dots) at $Re=170$.}
		\label{u1t_170}
	\end{center}
\end{figure}

Figure \ref{u1u2}(b) shows the projected orbit at $Re=180$.
After the first period-doubling bifurcation at $Re \simeq 179$,
the period-two orbit exhibits self-intersections when projected
onto a two-dimensional space spanned by any pair of variables.
As a result, the system state cannot be uniquely determined from
only two variables, since multiple states may correspond to the
same point in the projection. Therefore, more than two variables
are required to reconstruct the dynamics. In this case, three
variables are both necessary and sufficient. This conclusion is
supported by Takens' embedding theorem \cite{takens1981}, which
guarantees the reconstruction of smooth dynamical systems, and by
its extension, Embedology \cite{sauer1991embedology}, which
generalizes the result to chaotic attractors with fractal geometry.
According to these results, an embedding dimension of
$2\lfloor N_D \rfloor +1$ is sufficient for reconstruction,
except for a set of special cases, where $N_D$ denotes the
attractor dimension. For a periodic orbit, $N_D=1$, implying that
three variables are sufficient for an embedding.

\subsection{Chaotic case}

\begin{figure}%[htb]
	\begin{center}
		\psfrag{N}{$N$}	
		\psfrag{L2}{\rotatebox{180}{$L_2$}}
		\includegraphics[width=8cm,height=8cm]{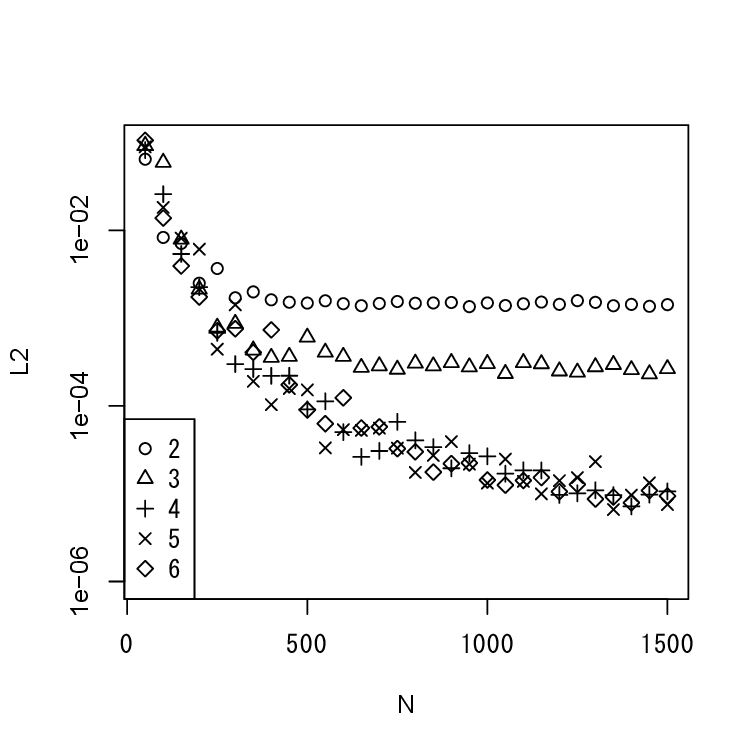}
		\psfrag{time}{$t$}
		\psfrag{enstro}{$W$}
		\includegraphics[width=8cm,height=8cm]{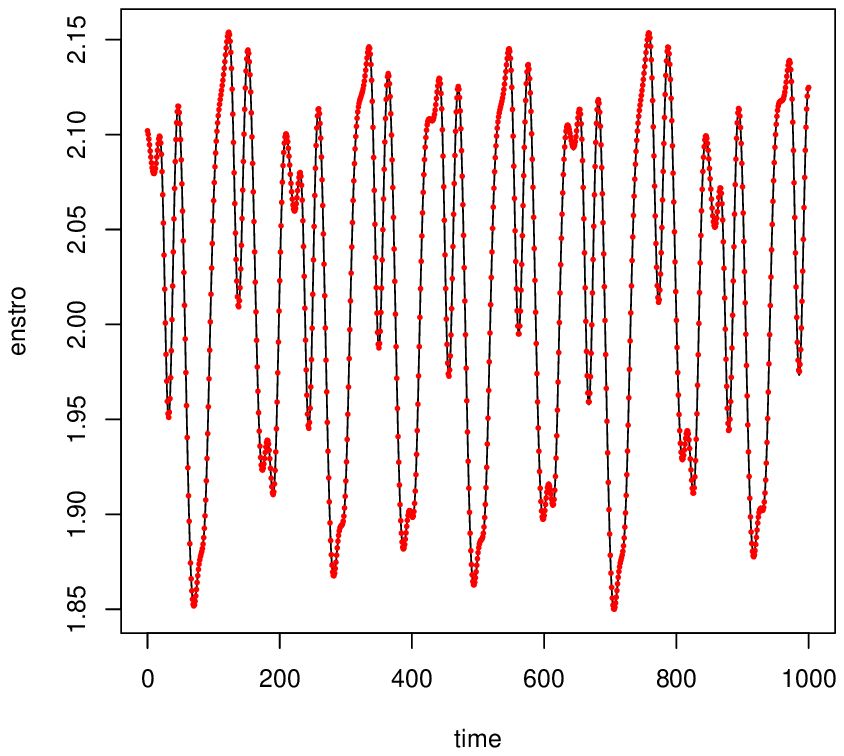}
		\caption{
			Same as figure \ref{u1t_170} but for $Re=183$.}
		\label{u1t_183}
	\end{center}
\end{figure}

After the onset of turbulence via a period doubling cascade, 
the orbit becomes chaotic as shown in Fig. \ref{u1u2} (c).
Figure \ref {u1t_183} (top) shows that when there are four or more 
variables, a low-dimensional system can be constructed where the errors 
converge quickly. 
However, using three variables leads to a significant error.
Because the dimension of the attractor is slightly larger than two, 
three variables are necessary for state determination and 
four variables are necessary for multivalency.
Five variables are sufficient for the construction of this attractor 
from the theoretical condition $2[N_D]+1$ \cite{sauer1991embedology}, 
where $[N_D]$ is slightly larger than 2 at the beginning of chaos. 
Even in this chaotic case, the orbit
predicted by a low-dimensional system constructed 
with a sufficient number of variables can trace accurately 
the orbit obtained from the DNS.
Figure \ref{u1t_183}(bottom) compares the time sequence of the constructed five-dimensional system using 1500 training samples with that of the Navier--Stokes system.

\begin{figure}%[htb]
	\begin{center}
		\psfrag{time}{$t$}
		\psfrag{enstro}{$W$}
		\includegraphics[width=8cm,height=8cm]{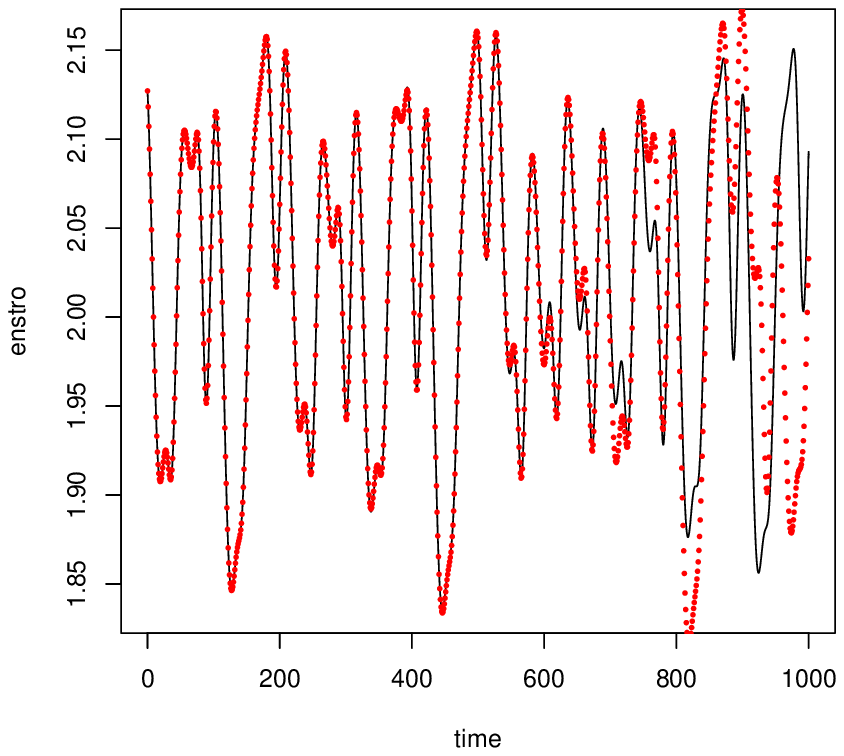}
		\psfrag{enstro}{$W$}
		\psfrag{pdf}{PDF}	
		\includegraphics[width=8cm,height=6cm]{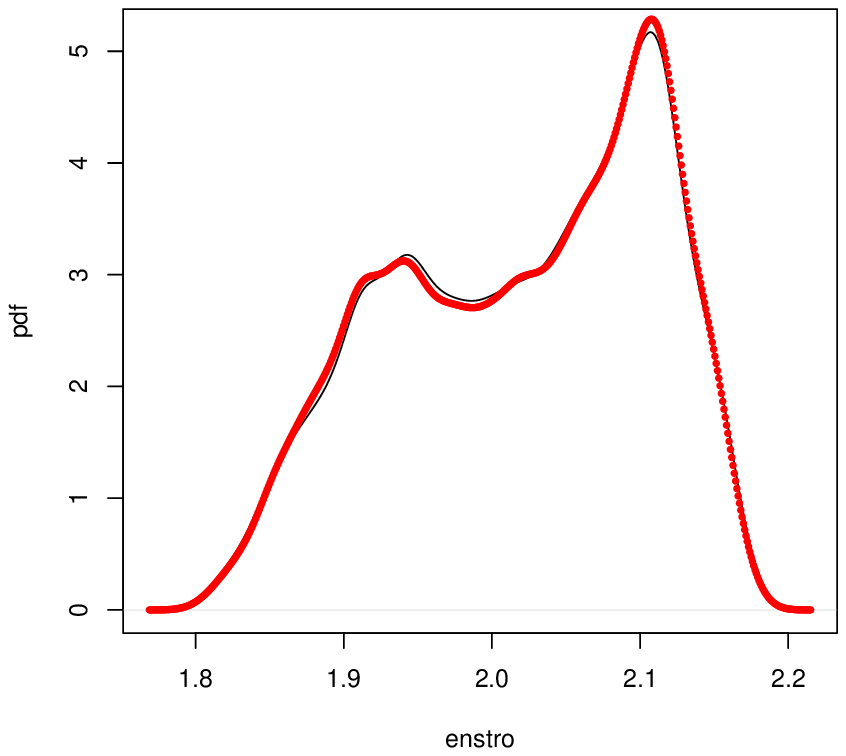}
		\caption{
			Comparison between the Navier--Stokes (solid line) and 
			the constructed low-dimensional system (dot) at $Re=183.1$. 
			(top) Trajectory.   
			(bottom) Probability density functions of $W$. } 
		\label{u1t_1831}
		\label{u1pdf0.10_1831}
	\end{center}
\end{figure}

At \(Re \simeq 183.02\), the chaotic attractor expands through an internal crisis. Figure \ref{u1u2}(d) shows its projection onto the \(E_y\)-\(E_z\) plane at \(Re=183.1\), immediately after the crisis. Again, a low-dimensional model can be constructed using four variables, even though fewer variables are sufficient to determine the system state (see Fig. \ref{u1t_1831}).
Since the turbulent state is orbitally unstable, it is impossible to track the Navier--Stokes trajectory beyond a time scale comparable to the inverse of the largest Lyapunov exponent. The three largest Lyapunov exponents, evaluated using the method of 
Shimada snd Nagashima (1979) \cite{shimada1979numerical}, are \(0.0062\), \(0.0000\), and \(-0.0190\) at \(Re=183.1\).
Although the system formally possesses two zero Lyapunov exponents associated with translational invariance in time \(t\) and in the streamwise direction \(x\), the latter is not taken into account because the attractor does not drift in the \(x\) direction. The system does not possess translational symmetry in the spanwise direction \(z\) because mirror symmetry about \(z=0\) is imposed.
The corresponding Lyapunov dimension \cite{kaplan1979chaotic} is \(D_{\rm L}\simeq 2.33\).

Figure \ref{u1pdf0.10_1831} (bottom) shows the probability density function (PDF) of \(W\) at \(Re=183.1\) for both the Navier--Stokes system and the reconstructed low-dimensional model. The PDFs are computed from trajectories of length \(T=100000\).
The PDF obtained from the low-dimensional model closely matches that of the full Navier--Stokes system, indicating that the essential statistical properties of the turbulent dynamics are retained in the reduced-order representation.

\subsection{Low-dimensional system including Reynolds number}

We next construct a low-dimensional model that explicitly incorporates \(Re\) as a parameter using a limited number of training samples. This is achieved by including \(Re\) among the arguments of \(F_i\) in equation (5). The resulting time-increment maps are constructed with \(M=8\).
For the model training, data are sampled at 25 Reynolds numbers from \(Re=164\) to \(188\) in increments of unity. For the steady states at \(Re=164\), \(165\), and \(166\), a single sample is used for each Reynolds number. For the periodic states in the range \(167 \leq Re \leq 182\), 50 samples are used for each Reynolds number. For \(Re \geq 183\), where the flow may exhibit turbulent dynamics, 100 samples are used for each Reynolds number.

\begin{figure}%[tb]
	\begin{center}
		\psfrag{re}{$Re$}
		\psfrag{sqrt_ecf}{$E_{cf}^{1/2}$}	
		\includegraphics[width=13cm]{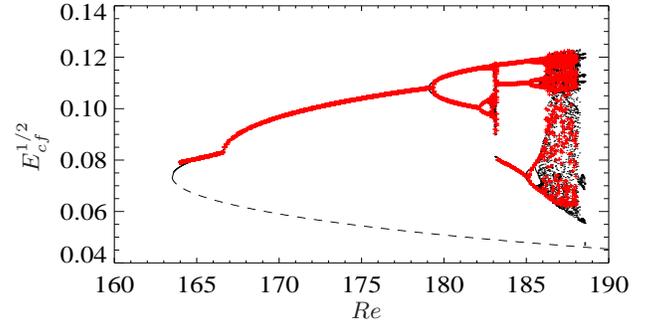}
		\caption{Bifurcation diagram of the constructed  low-dimensional 
			system (red symbol) and the Navier--Stokes system (black).}
		\label{plot_model2}
	\end{center}
\end{figure}

Figure \ref{plot_model2} shows the bifurcation structure of the reconstructed low-dimensional model. Only the asymptotic states (attractors) are plotted for both the DNS and the reduced system. For the latter, attractors are computed at intervals of \(Re=0.1\), an order of magnitude finer than the spacing of the training Reynolds numbers.
The agreement between the two bifurcation diagrams is excellent. In particular, the reduced model correctly captures bifurcations occurring between training points, demonstrating its capability to interpolate the underlying dynamics over the Reynolds-number range considered.

\section{Conclusion}

Using Kernel Quantile Regression \cite{takeuchi2006nonparametric}, we constructed low-dimensional map-based models that accurately predict low-Reynolds-number turbulent flows when the attractor dimension is small. Once a reduced-order model incorporating the control parameter is obtained, turbulent trajectories can be reproduced at a computational cost that is substantially lower than that of direct numerical simulation (DNS).
We also expressed physically important quantities as functions of the variables in the reduced system, providing insight into the relationship between the phase-space structure and the underlying flow physics.
In a preliminary study, the present approach was shown to be applicable, to some extent, to the same computational domain at \(Re=550\), which is slightly above the lower Reynolds-number limit for sustained turbulence. In that analysis, 32 measurements of the streamwise velocity in physical space were used, indicating that the required observables can be obtained experimentally.
Even for such a small domain and relatively low Reynolds number, however, the curse of dimensionality remains a major obstacle to the construction of highly accurate reduced-order models. For larger computational domains or higher Reynolds numbers, direct learning of the system dynamics becomes substantially more challenging, even when the attractor dimension is much smaller than that of the full DNS.
For such systems, it is essential to identify optimal latent variables that capture the dominant dynamics and enable deterministic or stochastic prediction of future states using only a small number of degrees of freedom. Recent studies suggest that autoencoder-based deep-learning techniques can discover compact latent representations of turbulent flows, providing promising feature variables for efficient reduced-order modeling \cite{vinuesa2024rom}.

\begin{acknowledgment}

This work was supported by the NIFS Collaboration Research program
(NIFS25KISC016), and JSPS KAKENHI Grant Number 24K07323.

\end{acknowledgment}

%\bibliographystyle{jpsj}  % jpsj.bst ���w��
%\bibliography{ref} % .bib �t�@�C�����i�g���q�Ȃ��j
\bibliography{ref1.2} %hoge.bib�����g���q���O�������O
\bibliographystyle{jpsj} %�Q�l�����o�̓X�^�C

\end{document}